\documentclass[twoside,12pt]{article}
\usepackage{epsfig}
\usepackage{amsmath}
\usepackage{amssymb}

\begin{document}

\title{Isospin relations for the tau decay modes}

\author{A. ROUG\'E\\
Laboratoire Leprince-Ringuet, \\
\'Ecole Polytechnique, \\ 
91128 Palaiseau, France\\ 
E-mail: rouge@in2p3.fr}


\maketitle

\begin{abstract}
Since the fifties, isospin relations have been used in particle physics to 
understand the properties of multihadrons final states. 
In the case  of the tau lepton, they allow to relate
the partial widths of the decay modes to the cross sections 
of $e^+e^-$ annihilations. 
A pedagogical introduction to the construction of isospin states 
for meson systems and an
updated review of the use of isospin relations in the study of the tau lepton 
are presented.
\end{abstract}

\begin{center}

{\bf Dedicated to Roberto Salmeron on the occasion of his 80th birthday.}
\end{center}

\section{Introduction}
In 1953, at the Bagn\`eres  de Bigorre conference~\cite{bdb}, which Roberto Salmeron 
attended as member of the Manchester group,  Dalitz showed the following 
inequality~\cite{dalitz}:
\begin{equation} \label{eq:dalitz}
1/4\leq\frac{\tau^+\to \pi^+\pi^0\pi^0}{\tau^+\to \pi^+\pi^+\pi^-}\leq 1\,.
\end{equation}
The $\tau^+$ in Eq.~\ref{eq:dalitz}, partner of the $\theta^+$ in the celebrated 
puzzle, is nowadays known  as $K^+$. The hypotheses leading to Eq.~\ref{eq:dalitz} 
were the existence 
of an isospin triplet ($\tau^+,\tau^0,\tau^-$) and the conservation of
isospin in the $\tau$ 
decay. None of them was founded. Nevertheless,  owing to the 
$|\,\Delta \vec I\,| =1/2$ rule, Eq.~\ref{eq:dalitz} survived the introduction of 
the Gell-Mann Nishijima 
scheme~\cite{gellmann,nishijima} and stimulated the discovery of the 
$K^+ \to \pi^+ \pi^0\pi^0$ decay mode~\cite{crussard}.

More than thirty years later, the same inequality was written in a 
paper~\cite{gilman} devoted to the ``calculation of exclusive decay
modes of the tau'', but the $\tau$ 
studied in the paper was the $\tau$ lepton. 
This coincidence can serve to illustrate the 
longevity and the generality of the isospin relations.   
However, in the following pages,  we will discuss only their applications~\cite{gilman,gilman2,clegg,sobie,ar1,ar2}
to the decay modes  of the $\tau$ lepton.

\section{Isospin relations in hadronic final states}

The  proof of  an equation like Eq.~\ref{eq:dalitz} starts from 
the identification of the different possible isospin states for the hadronic 
system. The amplitudes for a given charge-configuration  
($\pi^+\pi^+\pi^-$ and $\pi^+\pi^0\pi^0$ for three $\pi$'s  with $Q=1$) 
are linear combinations  of the isospin amplitudes;  
thus the partial widths are linear combinations of the squared amplitudes
and their interference terms.

Generally, a large fraction of the interference terms are killed by the 
integration over the phase-space. 
The remaining terms are bounded by the Schwartz inequality~\cite{michel}. 
The resulting constraints can be geometrically represented by an allowed 
 convex domain 
in the space of the charge-configuration fractions, 
$f_{cc}=\Gamma_{cc}/\Gamma$. 
If  all the interferences vanish,  the domain is a polyhedron,
 convex hull of the points that describe each isospin state.

Hence, the first step in the construction of the allowed domain  
is the setting up of a basis for the isospin states, adapted to the
implementation of the Pauli principle.

\section{Isospin states of $n\,\pi$ systems}
 \label{sec:iso} 
Such a basis was constructed by Pais~\cite{pais} with the object of studying 
the many pion systems produced in $\bar p p$ and $\bar p n$ annihilations.
The construction is based on two simple remarks\,: {\it i)} the representations of 
the isospin group SU(2)  relevant for $n\pi$ systems are also representations
of SO(3), {\it ii)} the group of $3\times 3$ orthogonal unimodular matrices, SO(3), is a subgroup of the group of $3\times 3$ unimodular matrices, SL(3).  
Thus, if $V$ is the three-dimensional space of the isospin states  of one pion,
the space of the states of $n$ pions,
 $V^{\otimes n}$, supports representations of both SL(3) and the symmetric group
 ${\mathcal S}_n$, which acts on $V^{\otimes n}$ by permuting its factors.
Standard properties of the representations of linear groups~\cite{weyl,boerner,sternberg}
imply the decomposition
\begin{equation}\label{eq:sunsp} V^{\otimes n}=\bigoplus^\lambda E_\lambda\otimes F_\lambda,
\end{equation}
where the symbol\footnote{Also denoted~\cite{pais,boerner} ``symmetry class''.} $\lambda=(\lambda_1,\lambda_2,\lambda_3)$, with $\lambda_1\ge 
\lambda_2\ge \lambda_3$ and $\lambda_1+\lambda_2+\lambda_3=n$,
is associated to a three-row Young diagram; $F_\lambda$ is the irreducible 
representation of 
 ${\mathcal S}_n$ determined 
by the diagram, and $E_\lambda$  an irreducible representation of SL(3).
For a given $n$, the correspondence between the representation of SL(3) and 
$\lambda$ is one--one. The representation $E_\lambda$ is 
characterized by  the two numbers: $\lambda_1-\lambda_2$ and $\lambda_2-\lambda_3$.

As a representation of SO(3), $E_\lambda$ is no longer irreducible because 
of the invariance under SO(3) of the contraction operation~\cite{weyl,racah}. 
Its decomposition
into irreducible representations of SO(3) reads
\begin{equation}
\label{eq:slso}
 E_\lambda=\bigoplus^IN_I(\lambda)D_I,
\end{equation}
where $D_I$ is the $(2I+1)$-dimensional irreducible representation (integer isospin I). 
The multiplicity $N_I(\lambda)$ was computed by Racah~\cite{racah}. It can be written:
\begin{equation}
\label{eq:racah}
N_I(\lambda)=
\phi(\lambda_1-\lambda_3-I+2)
-\phi(\lambda_2-\lambda_3-I+1)
-\phi(\lambda_1-\lambda_2-I+1),
\end{equation}
where $\phi(x)$ is the greatest integer contained in  $x/2$ for $x> 0$, 
and $0$ for $x\le 0$.
For systems of two  pions, we get:
\begin{center}
\begin{tabular}{ccccc}
&$\lambda$&$N_0$&$N_1$&$N_2$\\
\raisebox{-12pt}{\epsfxsize=20pt\epsfbox{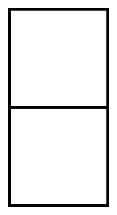}}&(1,1,0)&0&1&0\\[10pt]
\raisebox{-7pt}{\epsfxsize=30pt\epsfbox{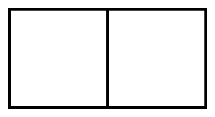}}&(2,0,0)&1&0&1\\
\end{tabular}
\end{center}
and, for three pions, 
\begin{center}
\begin{tabular}{cccccc}
&$\lambda$&$N_0$&$N_1$&$N_2$&$N_3$\\
\raisebox{-17pt}{\epsfxsize=20pt\epsfbox{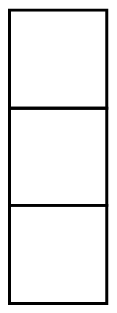}}&(1,1,1)&1&0&0&0\\[10pt]
\raisebox{-12pt}{\epsfxsize=30pt\epsfbox{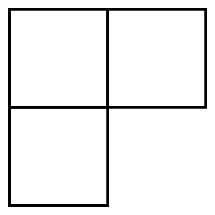}}&(2,1,0)&0&1&1&0\\[10pt]
\raisebox{-7pt}{\epsfxsize=40pt\epsfbox{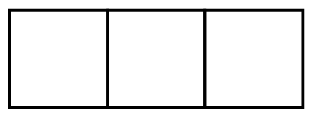}}&(3,0,0)&0&1&0&1\\
\end{tabular}
\end{center}
For any $\lambda$, Eq.~\ref{eq:racah} implies  the relation
\begin{equation}\label{eq:01} 
N_0(\lambda)+N_1(\lambda)=1,
\end{equation} 
which can also be obtained directly from the symmetry 
($\square\hspace{-2pt}\square$) of a contraction.
If both $\lambda_1-\lambda_2$ and $\lambda_2-\lambda_3$ are even, $N_0=1$,
 otherwise $N_1=1$. 

From the equations (\ref{eq:sunsp}), (\ref{eq:slso}) and (\ref{eq:01}),
we see that the sole degeneracy of the  $n$-$\pi$ states $|I_3,\lambda\rangle$ 
with $I\le 1$, ($I_3$ and $\lambda$ fixed)
   is due to the permutation symmetry. These states form an 
irreducible representation of ${\mathcal S}_n$. Since a permutation preserves
the charge-configuration ($n^+,n^0,n^-$) of a $n$-$\pi$ system, the 
irreducibility implies that the Clebsch-like coefficients used to write the isospin states 
as combinations of charge-configuration states are  determined
by $I_3$ and $\lambda$ only\footnote{The elements of the group algebra used to build an
orthogonal basis of the representation  give also orthogonal charge configuration states~\cite{pais}.}. In other words~\cite{pais}, the coefficients of the 
charge-configurations are a class property.  

Furthermore, since the permutation symmetry properties of the momentum and isospin 
amplitudes are the same, because of the Pauli principle, integrating over the 
phase-space kills all the interference terms for a $n$-$\pi$ system with 
$I\le 1$; the 
allowed domain in the space of the charge-configuration fractions is a 
polyhedron.

Let's take the simple example of the $Q=1$, $I=1$ three-$\pi$ system 
alluded to in the introduction. 
For SO(3), an isospin one is a vector,  and any vector made of 
three vectors can be written 
\[\alpha\,(\vec b\cdot\vec c\,)\,\vec a+\beta\,(\vec c\cdot\vec a\,)\,\vec b
+\gamma\,(\vec a\cdot\vec b\,)\,\vec c.\]

As an isospin function, $(\vec a\cdot\vec b\,)\,\vec c$
describes an $I=1$ state with the two first $\pi$ in an $I=0$ state: 
$(\pi^+\pi^--\pi^0\pi^0+\pi^-\pi^+)\pi^+$. 
It is then straightforward to write the other terms by cyclic permutations and
get the ratio
\begin{equation}
R=\frac{\pi^+\pi^0\pi^0}{\pi^+\pi^+\pi^-}=\frac{|\alpha|^2+|\beta|^2+|\gamma|^2}
{|\beta+\gamma|^2+|\alpha+\gamma|^2+|\alpha+\beta|^2}.
\end{equation}
The symmetry class $\lambda=(3,0,0)$, is associated to  the one-dimensional
space of completely symmetric states ($\alpha=\beta=\gamma$)
for which the ratio is $R=1/4$; the class
$\lambda=(2,1,0)$ to the two-dimensional orthogonal space 
($\alpha+\beta+\gamma=0$), with $R=1$. 
An example of state in the  $(2,1,0)$ class is given by a $\rho\pi$ system,
which is
represented by
$(\vec a\wedge\vec b\,)\wedge\vec c$, where the vector product ($\wedge$)
is interpreted as the combination of two isospins one into an isospin one.

Thus the charge-configuration
fractions are:
\begin{equation}
f_{\pi^+\pi^0\pi^0}=\frac{1}{2}W_{(210)}+\frac{1}{5}W_{(300)},\hspace{2em}
f_{\pi^+\pi^+\pi^-}=\frac{1}{2}W_{(210)}+\frac{4}{5}W_{(300)},\label{eq:3pi}
\end{equation}
with $W_{(210)}+W_{(300)}=1$.

The weights $W_\lambda$ depend  on the dynamics.
In the Fermi statistical model~\cite{fermi,cerulus}, they
 are  proportional to the dimension of the representation $F_\lambda$
of the permutation group ${\mathcal S}_n$; here   $W^{stat}_{(210)}=2/3$ and
$W^{stat}_{(300)}=1/3$.

If $I=2$ states are allowed, they have to share the symmetry properties
of $I=0$ or $I=1$ states, so that interference terms must be taken 
into account. It can be checked on Eq.~\ref{eq:racah} that, for $n<6$, 
$N_2(\lambda)\leq 1$ for all $\lambda$. Thus, for $n<6$ and $I\leq 2$,
the charge-configuration coefficients are determined by $I$, $I_3$ and 
$\lambda$  only.

Tables of the coefficients can be found in the literature~\cite{pais,pilkuhn}.
They are computed by explicitly constructing tensors with 
the required symmetry,  as in the previous example, or by more sophisticated 
methods~\cite{pais,pilkuhn,chacon}.

\section{Semileptonic decays of the $\tau$ lepton}       
The possible hadronic systems\footnote{The $\eta$ channels have to be 
treated separately because $\eta$ decay violates isospin.} ($h$) in the semileptonic decay $\tau\to\nu h$
are: $n\pi$, $\eta n\pi$, $K n\pi$, $K\eta n\pi$, and $K\bar K n\pi$.

The properties of the charged weak current imply that the total isospin is 1
for the strangeness zero final states, and 1/2 for the others. Thus the isospin
of the $n\pi$ system is 1 for $h=n\pi$ and $h=\eta n\pi$; 0 or 1 for 
$h=K n\pi$; 0, 1 or 2 for $h=K\bar K n\pi$. 

For all the cases but 
$h=K\bar K n\pi$ the
isospin amplitudes can be labelled by the symmetry class $\lambda$ only and the
interference terms are killed by the integration over phase-space. Thus the 
partial width for a given charge-configuration ($cc$) can be written:
\begin{equation}
\Gamma^{cc}=\sum\nolimits_\lambda\, C^{cc}_{\lambda}\, \Gamma^\lambda,
\end{equation}
where the coefficients $ C^{cc}_{\lambda}$ can be found in tables~\cite{ar1,pais}.

For positive G-parity systems \big($h=2n\pi$, $h=\eta(2n+1)\pi$\big), 
the weak current is related by an isospin rotation to the electromagnetic 
current, hence relations can be established between the $\tau$ partial widths 
and the cross sections of $e^+e^-$ annihilations.

For  $h=K\bar K n\pi$, a more detailed analysis taking into account the interferences
is needed.
\begin{figure}[ht]
\centerline{\epsfxsize=7.5cm\epsfbox{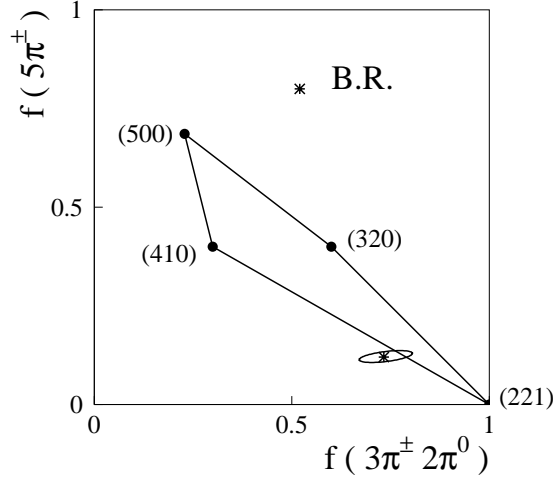}}   
\caption{\label{fig:5pi}The decay fractions for $\tau\to\nu 5\pi$. The experimental point
($\eta$ subtracted) is shown with the $1\sigma$ (39\% probability) contour.}
\end{figure}

 \subsection{$\tau\to\nu(2n+1)\pi$}
For a $3\pi$  system, there are two possible symmetry classes and
the isospin constraint is Eq.~\ref{eq:dalitz}. 

Experimentally, the ratio 
$\pi^-\pi^-\pi^+/\pi^-\pi^0\pi^0$ is nearly 1 because of the dominance of the 
$\rho\pi$ intermediate state. A detailed analysis of the final 
state~\cite{cleoa1}, taking into account the isospin symmetry breaking
 caused by the  difference of the  $\pi^0$ and $\pi^\pm$ masses, 
predicts a ratio 0.985, in good 
agreement with the measurements~\cite{pdg}.

For a $5\pi$ final state, three charge-configurations and four symmetry classes
are present. 
The production of $\eta$ contributes to the $2\pi^-\pi^+2\pi^0$
and $\pi^- 4\pi^0$ final states.

The comparison of the measured branching ratios~\cite{pdg}
(after $\eta$ subtraction) with the allowed domain is made in 
Fig.~\ref{fig:5pi}. It shows the dominance of $\lambda=(2,2,1)$,
which is due to the $\omega\pi^-\pi^0$ intermediate state. 
\subsection{$\tau\to\nu 2n\pi$}
The final states of an even number of pions are produced by the vector
current, which is related by an isospin rotation to the electromagnetic 
current. Since the isospin rotation commutes with the permutations,
the relation between $\tau$ partial widths and $e^+e^-$ 
cross sections~\cite{gilman} can be written for each symmetry class $\lambda$: 
\begin{equation}\label{eq:cvc}
\frac{1}{\Gamma_{\nu\bar\nu e}}\frac
{d\Gamma^{\lambda}_{\nu 2n\pi}}{dm^2}=\frac{3\cos\theta_c^2}
{2\pi\alpha^2m_{\tau}^8}m^2(m_{\tau}^2-m^2)^2(m_{\tau}^2+2m^2)
\sigma^{\lambda}_{e^+e^-\to 2n\pi}(m^2).
\end{equation}
\begin{figure}[ht]
\centerline{\epsfxsize=7.5cm\epsfbox{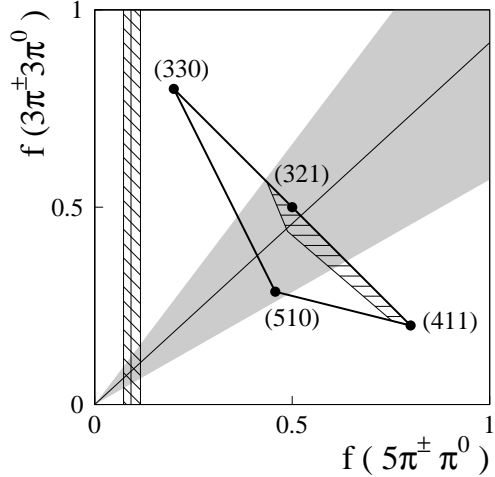}}   
\caption{The decay fractions for $\tau\to\nu 6\pi$. The 
grey region corresponds to one standard deviation from the measured
$B_{3\pi^\pm 3\pi^0}/B_{5\pi^\pm\pi^0}$ ($\eta$ subtracted). 
The hatched regions are estimations ($1\sigma$) 
from $e^+e^-$ annihilations data:
ratio $\sigma_{e^+e^-\to 6\pi^\pm}/\sigma_{e^+e^-\to 4\pi^\pm 2\pi^0}$ 
for the diagonal band; comparison of  $B_{5\pi^\pm\pi^0}$ and the total cross section $\sigma_{e^+e^-\to( 6\pi)^-}$ for the vertical band.
\label{fig:6pi}}
\end{figure}
There is only one class for  $2\pi$ systems. For $4\pi$ final states
the possible classes are $\lambda=(3,1,0)$ and $\lambda=(2,1,1)$.
Since there are two charge-configurations for both $\tau$ decays and
$e^+e^-$ annihilations, the correspondence between cross sections
and partial widths is very simple:
\begin{eqnarray*}
\sigma_{2\pi^+2\pi^-}&~\longleftrightarrow~&
2\Gamma_{\pi^-3\pi^0}\\
\sigma_{\pi^+\pi^-2\pi^0}&\longleftrightarrow&
\Gamma_{\pi^-\pi^-\pi^+\pi^0}-\Gamma_{\pi^-3\pi^0}.
\end{eqnarray*}  
Thorough comparisons~\cite{davier} of $\tau$-decay and $e^+e^-$-annihilation 
data for the two- and four-pion channels, including the consideration 
of isospin symmetry breaking, have been made recently in order to improve 
the theoretical determination of the muon anomalous magnetic moment $a_\mu$.
They show some discrepancies between the $e^+e^-$ and $\tau$ data, 
as well as between different $e^+e^-$ experiments.

Four classes can contribute to the $6\pi$ states production, and only three
charge-con\-fi\-gu\-ra\-tions are possible in $\tau$ decays as well as in $e^+e^-$ 
annihilations. Therefore, even with complete measurements it would not be 
possible to  predict the partial widths from the cross sections or conversely.
Nevertheless, with two measurements in $e^+e^-$ annihilations 
($\sigma_{3\pi^+3\pi^-}$, $\sigma_{2\pi^+2\pi^-2\pi^0}$) and in $\tau$ decays
($B_{3\pi^-2\pi^+\pi^0}$, $B_{2\pi^-\pi^+3\pi^0}$), it is possible to determine
the contributions of the four classes, or, at least, to check the consistency 
of the different measurements.

Figure~\ref{fig:6pi} displays the allowed region in the plane of the 
charge-configuration fractions and the ratio 
$B_{2\pi^-\pi^+3\pi^0}/B_{3\pi^-2\pi^+\pi^0}$ of the measurements~\cite{cleo6},
after $\eta$ subtraction (grey area).

The large $\sigma_{2\pi^-2\pi^+2\pi^0}/\sigma_{3\pi^-3\pi^+}$ ratio observed
in $e^+e^-$ annihilations implies the dominance of $\lambda=(3,2,1)$
and/or  $\lambda=(4,1,1)$, quantitatively shown~\cite{ar1} by the diagonal hatched 
region in Fig.~\ref{fig:6pi}.
The location of the intersection of the two regions corresponds to 
$\lambda=(3,2,1)$, in agreement with the observed~\cite{cleo6} importance of the
\begin{figure}[ht]
\centerline{\epsfxsize=7.5cm\epsfbox{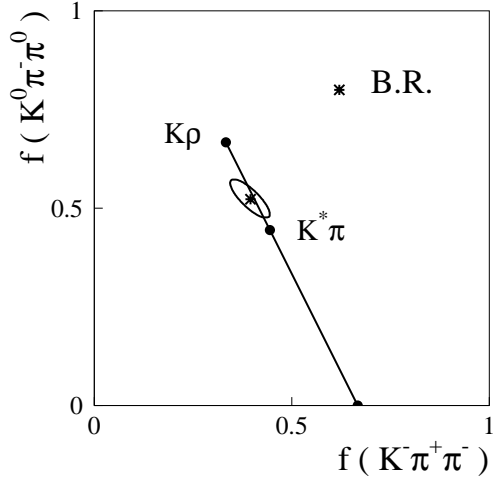}}   
\caption{The decay fractions for $\tau\to\nu K\pi\pi$. The experimental point
 is shown with the $1\sigma$ (39\% probability) contour.
The solid line is the isospin constraint (Eq.~\ref{eq:k2pi}). \label{fig:kpipi}}
\end{figure}
$\omega$ production.

Discrepancies appear when the total cross section for  $e^+e^-$ annihilations
into $6\pi$ is taken into account. It allows an estimation~\cite{eidelman}
of the total $B_{6\pi}$ branching ratio, which together with the measured 
$B_{3\pi^-2\pi^+\pi^0}$ gives the vertical hatched band in Fig.~\ref{fig:6pi},
clearly incompatible with the other estimations.

Two hypotheses can be contemplated: either the  $e^+e^-$ cross sections are 
overestimated by a factor of roughly four, or the $e^+e^-$ annihilations
into $6\pi$ receive a large contribution from $I=0$, $\eta3\pi$ final states.
In the second hypothesis, the cross section for \mbox{$e^+e^-\to \pi^+\pi^-+ neutrals$}
would be three times larger than $\sigma_{2\pi^-2\pi^+2\pi^0}$.

The contribution of the $6\pi$ channel to the estimation of $a_\mu$ is 
small~\cite{davier}, however the second hypothesis, if true, could have a not
 completely negligible impact on the estimation.

\subsection{$\tau\to\nu  K n\pi$}

For $\tau\to\nu K\pi$, the isospin 1/2 implies the ratio
$K^-\pi^0/\bar K^0\pi^-=1/2$,  to be compared with the experimental 
value~\cite{pdg}\[ B_{K^-\pi^0}/B_{\bar K^0\pi^-}=0.51\pm 0.04\,.\]

For $K n\pi$ systems, the number of charge-configurations is greater than 
the number of symmetry classes. The resulting  relations between 
branching ratios are
\begin{eqnarray}\label{eq:k2pi}
B_{K^-\pi^+\pi^-}&=&\frac{1}{2}B_{\bar K^0\pi^-\pi^0}+2B_{K^-\pi^0\pi^0}\\
B_{\bar K^0\pi^-\pi^-\pi^+}&=&B_{\bar K^0\pi^-\pi^0\pi^0}+2B_{K^-\pi^0\pi^0\pi^0}
\label{eq:k3pi}\end{eqnarray}
for three and four hadron final states.
The data for the $K2\pi$ final states are shown in Fig.~\ref{fig:kpipi}.
The location of the experimental point is consistent with the observed
dominance of the intermediate states $K^*\pi$ and $K\rho$.  
\subsection{$\tau\to\nu  K\bar K n\pi$}
\begin{figure}[ht]
\centerline{\epsfxsize=7.5cm\epsfbox{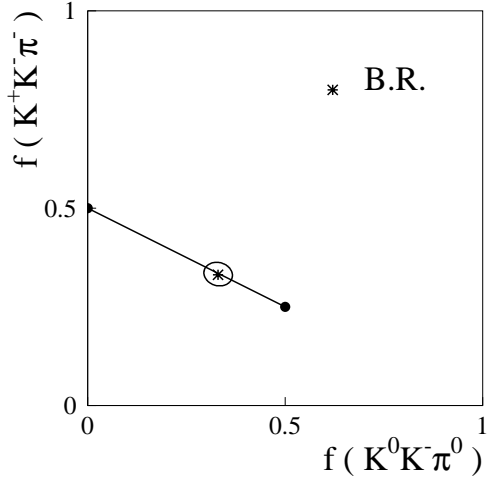}}   
\caption{The decay fractions for $\tau\to\nu K\bar K\pi$. The experimental point
 is shown with the $1\sigma$ (39\% probability) contour.
The solid line is the isospin constraint (Eq.~\ref{eq:kkpi}). \label{fig:kkpi}}
\end{figure}

In a $K\bar K$ system, the possible values of the $K\bar K$ isospin are 
0 and 1. For $I_{K\bar K}=0$ the isospin of the $n\pi$ system is $I_{n\pi}=1$;
it can be 0, 1 or 2 for  $I_{K\bar K}=1$.
Both axial  and vector currents contribute to the decay, leading to $G=+1$ (V)
and $G=-1$ (A) for the G-parity of the hadronic system, and 
$G_{K\bar K}=(-1)^nG$. 

Since the $J^P$ quantum numbers are different for the axial and vector 
currents, there is no V-A interference term in the partial widths.

The G-parity of a $K\bar K$ system is related 
to its isospin and orbital momentum by 
$G_{K\bar K}=(-1)^{I_{K\bar K}+l_{K\bar K}}$, thus for a given (V or A)
current the values of   $l_{K\bar K}$ associated to  $I_{K\bar K}=0$ and 
 $I_{K\bar K}=1$ are different and there is no 
  interference term between $I_{K\bar K}=0$ and $I_{K\bar K}=1$ 
amplitudes. 

This imply~\cite{ar1},
for any charge-configuration ($cc$) of the $n\pi$ system, the equality.
\begin{equation}\label{eq:kkpi}
\Gamma_{K^0\bar K^0(n\pi,cc)}=\Gamma_{K^+ K^-(n\pi,cc)}.
\end{equation} 
From CPT invariance, $\Gamma_{K_S K_S(n\pi,cc)}=\Gamma_{K_L K_L(n\pi,cc)}$, 
but the ratio $\Gamma_{K_S K_L(n\pi,cc)}/\Gamma_{K_S K_S(n\pi,cc)}$
is a free parameter depending on dynamics and the respective contributions of 
V and A currents~\cite{ar1}.

Figure~\ref{fig:kkpi} shows the agreement of the data~\cite{pdg,liu} with
Eq.~\ref{eq:kkpi} in the case of the $K\bar K\pi$ final state.
The ratio  $K^+K^-\pi^+/K^0K^-\pi^0$ is found equal to 1 in agreement with 
the observed dominance of the intermediate state $K^*\pi$.

The isospin amplitudes for a $K\bar K n\pi$  final state can be labelled by  
$I_{K\bar K}$, $I_{n\pi}$ and the symmetry class  $\lambda$ of the $n\pi$ system. 

If no  $I_{n\pi}=2$ amplitude is associated with $\lambda$, there is no possible 
interference term and the class is described by a point in the space of the
charge-configuration fractions. 

If $I_{n\pi}=2$ is possible, there is one interference term, but only one 
since $m_\tau<2m_K+6m_\pi$ (section \ref{sec:iso}), thus the class is 
described by a two-dimensional elliptic domain~\cite{ar2}.

\begin{figure}[ht]
\centerline{\epsfxsize=7.5cm\epsfbox{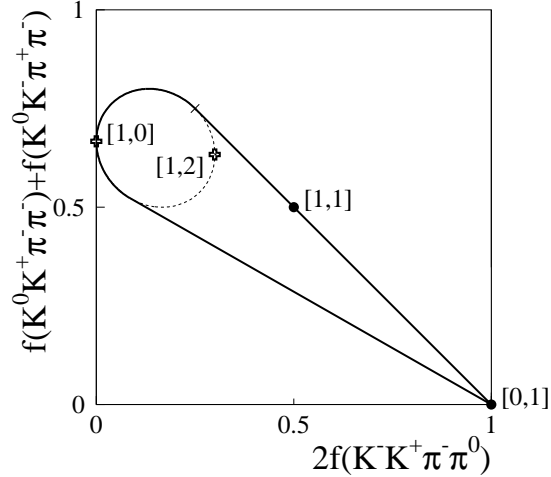}}   
\caption{Projection of the allowed domain for $\tau\to\nu K\bar K\pi\pi$.
The classes of amplitudes are labelled by the isospin values,
$[I_{K\bar K},I_{\pi\pi}]$. 
\label{fig:kkpipi}}
\end{figure}

The allowed domain in the space of the charge-configuration fractions is 
the convex hull of the points and ellipses associated with the symmetry classes
of the $n\pi$ system. Fig.~\ref{fig:kkpipi} shows a projection of this 
multidimensional domain in the case of $n=2$.
It implies the following inequality~\cite{ar2}:
\begin{equation}
B_{K^0K^-\pi^0\pi^0}\leq\frac{3}{4}(B_{\bar K^0K^+\pi^-\pi^-}
+B_{K^0K^-\pi^+\pi^-}),
\end{equation}
which, together with the relation 
$B_{\bar K^0\pi^-\pi^0\pi^0}\leq B_{\bar K^0\pi^-\pi^-\pi^+}$ (Eq.~\ref{eq:k3pi}),
gives the constraint 
\begin{equation}
B_{K_Sh^-\pi^0\pi^0}< B_{K_Sh^-h^-h^+}
\end{equation}
on topological branching ratios.
\section{Conclusion}
Isospin relations have been used in $\tau$ lepton physics for nearly 
twenty years.
Their first applications~\cite{gilman,gilman2} were the estimation of $\tau$ 
branching ratios from $e^+e^-$ annihilation data, the bounding of the
contributions of unobserved channels, and the elucidation of the 
``one prong problem''~\cite{bs,ws}. 

Today, a large number of decay modes are tabulated~\cite{pdg}, the order of magnitude of the smallest measured branching ratios 
is $10^{-4}$,
and the data from $\tau$ decays are used~\cite{davier} to complement and correct
the information given by the    $e^+e^-$ annihilations.

The observed discrepancies can only be solved by experiment.
However, quoting from Blackett's closing remark at the Bagn\`eres 
conference~\cite{blackett}:
``if the history of scientific discovery is any guide, the same increase 
of accuracy which will serve to settle our present controversies will equally, 
surely, itself bring to birth new controversies by leading of some discoveries.''

\end{document}